\documentclass[]{article}

\usepackage[margin=2.5cm]{geometry}
\usepackage{cite}

\usepackage[caption=false]{subfig}
\usepackage{lipsum}  
\usepackage[normalem]{ulem}
\usepackage{hyperref}
\usepackage{siunitx}
\usepackage{graphicx}
\graphicspath{{./figs/}}
\usepackage{amsmath,amssymb}
\usepackage{xcolor}
\usepackage{authblk}

\hypersetup{
	colorlinks = true,
	urlcolor   = blue,
	citecolor  = black,
	linkcolor = black
}

\newcommand{\Wi}{\mathrm{Wi}}
\newcommand{\De}{\mathrm{De}}
\renewcommand{\Re}{\mathrm{Re}}

\title{Splitting of localized disturbances in viscoelastic channel flow}

\author[1,2]{Ron Shnapp}

\author[1,3]{Victor Steinberg}

\affil[1]{Department of Physics of Complex Systems, Weizmann Institute of Science, Rehovot 76100, Israel}

\affil[2]{Swiss Federal Institute of Forest, Snow and Landscape Research WSL, Birmensdorf, Switzerland}

\affil[3]{The Racah Institute of Physics, Hebrew University of Jerusalem, Jerusalem 91904, Israel}

\date{}

\begin{document}

\maketitle

\begin{abstract}
	We examine the response of an inertia-less viscoelastic channel flow to localized perturbations. A simplified model shows that the non-linear interaction between the velocity and elastic stress fields can split an initial pulsed disturbance into two separate pulses if the initial disturbance and the base elastic stress are sufficiently high. In accordance, we demonstrate that a transition to a pulse-splitting regime can be achieved experimentally. These results suggest a possible new direction for studying the elastic instability of viscoelastic channel flows at high elasticity through the growth of localized perturbations.
\end{abstract}

% ==========================================================
%                          Introduction:
% ==========================================================

\section{Introduction}

Fluids supplemented by polymer chains exhibit a viscoelastic behavior, expressed by the appearance of internal elastic stresses. The elastic stress field is coupled to the fluid's motion due to polymer stretching by the velocity gradients~\cite{Larson1999, Smith1999, Liu2010}, as characterized by the Weissenberg number $\Wi$ ($\Wi=\frac{U}{L}\lambda$, where $U$, $L$ and $\lambda$ are the characteristic velocity, length scale, and longest polymer relaxation time, respectively). As a result, flow instabilities and elastic turbulence might occur in viscoelastic fluid flows at $\Wi\gg1$, even at vanishingly small Reynolds numbers, $\Re\ll1$ ($\Re=\frac{U\,L}{\nu}$, where $\nu$, is the fluid's kinematic viscosity)~\cite{Larson1992, Pakdel1996, Shaqfeh1996, Steinberg2021}. Despite several decades of research and the prevalence of viscoelastic fluids in industrial and biological processes, the understanding of instability mechanisms in the high elasticity regime ($\Re\ll1$ and $\Wi\gg 1$) is still lacking in several fundamental cases, such as in the channel flow (aka, plane Poiseuille flow)~\cite{Morozov2007, Steinberg2021} which is discussed here.

According to the theory of hydrodynamic instability~\cite{Drazin2004}, infinitely small perturbations to the viscoelastic plane Poiseuille flow should decay exponentially~\cite{Morozov2007}. However, two different theories predicted that these flows might be unstable due to either a sub-critical normal-mode instability~\cite{Meulenbroek2004, Morozov2005}, or due to a non-modal instability which leads to an algebraic transient growth of finite perturbations~\cite{Schmid2007, Jovanovic2010, Jovanovic2011}. Therefore, it is challenging to analyze the transition in these flows because finite perturbations are needed to provoke their instability. Accordingly, Refs.~\cite{Bonn2011, Pan2013, Qin2017, Qin2019} showed experimentally that strong flow perturbations above the instability lead to a chaotic, mixing-flow regime downstream of the perturbation location instead of a single fastest exponentially growing mode in the normal mode instability. More recently, experiments from our group~\cite{Jha2020, Jha2021} have characterized the transition to chaotic flow regimes and the structure of the flow following the transition. The transition was seen to agree with the non-modal instability scenario, and the instability was energized at low frequencies by elastic waves through an elastic analog of the Kelvin Helmholtz instability~\cite{Jha2021}. Furthermore, the present authors have shown~\cite{Shnapp2021} that the elastic instability can develop even due to very weak perturbations to the flow. Nevertheless, how the critical amplitude for the transition depends on $\Wi$ in the $\Re\ll1$ regime was not examined in experiments previously.

In the experimental studies mentioned above~\cite{Bonn2011, Pan2013, Qin2017, Qin2019, Jha2020, Jha2021, Shnapp2021}, the disturbance to the flow was achieved by altering the boundary conditions either at the inlet or at the center of the channel while characterizing the random flow that develops downstream from these locations. This approach may be termed the \textit{constant perturbation method} of characterizing the instability and transition~\cite{Barkley2016}, since the disturbance is always present in the apparatus. In contrast, an alternative method for studying flow instabilities is by producing a \textit{localized disturbance} and tracking the perturbed fluid packet to observe whether it decays or grows with time. This approach, which may be termed the \textit{localized perturbation method}~\cite{Barkley2016}, was used successfully to characterize the transition to inertial turbulence in the revolutionary works of Refs.~\cite{Wygnanski1973, Faisst2004, Darbyshire1995, Hof2003, Avila2011} (see Fig.2 in Ref.~\cite{Barkley2016} for a illustration of the two methods). The \textit{localized perturbation method} was never attempted in low Reynolds number viscoelastic shear flows in the past, despite its virtues in studying systems with a non-modal instability~\cite{Faisst2004} such as the flow examined here. Furthermore, the localized perturbation method is advantageous for studying the dependence of the critical value of the transition on the amplitude of the perturbations since the latter can be changed throughout the experiment (i.e., as performed in Ref.~\cite{Hof2003} for inertial turbulence).

This work presents the first analysis of the viscoelastic channel flow instability in the high-elasticity regime through the localized perturbation method. We begin by using a simplified model to demonstrate that elastic stresses in the fluid may cause single-pulsed perturbations to split into two pulses. Following that, we demonstrate a transition to a pulse-splitting regime experimentally. Specifically, we show that for low Weissenberg number values, $\Wi<\Wi_c$, initial pulse perturbations persist without splitting for very long times (up to several $\lambda$); however, for higher $\Wi$, the initially localized pulse may undergo a series of seemingly random splittings. Furthermore, we demonstrate that the critical value for the transition depends on the strength of the perturbation. These results suggest a new research direction by showing a plausible scenario for the growth of perturbations through the localized perturbation approach.

% ================================================================
%                            The model                              
% ================================================================

\section{An elastic pulse-splitting process}

\begin{figure}
	\centering
	\includegraphics[width=7.5cm]{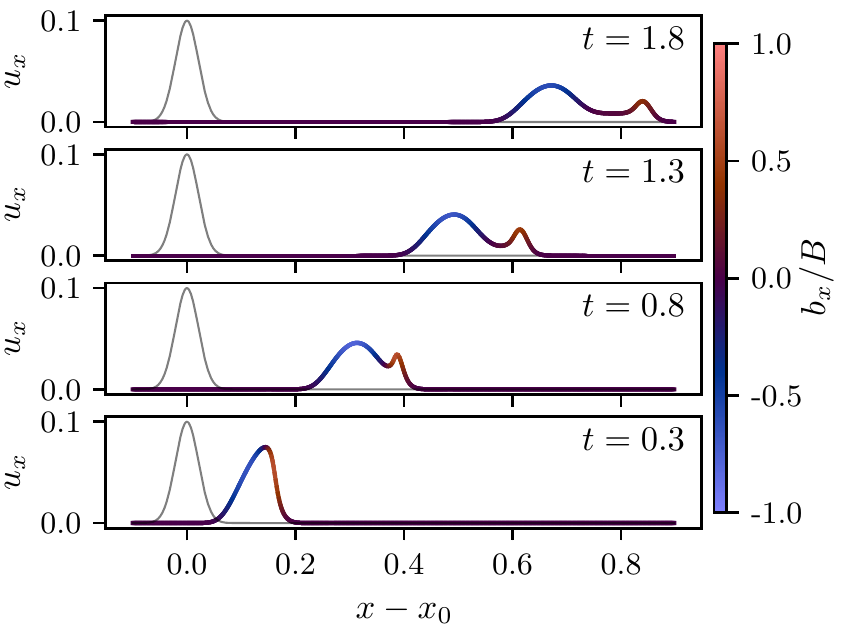}
	\caption{Results of the one dimensional model, showing the streamwise velocity as a function of position at different times; the modulus of the elastic stress is shown as the shade of the curves. The initial conditions used are a Gaussian pulse for the velocity $u_x = 0.1 \exp \left[-(x/0.03)^2 \right]$ and a zero stress $b_x(x)=0$. The parameters used are $B=-0.05$, $U=0.4$, $\lambda=2.0$, and $\nu=10^{-4}$. \label{fig:1d_model}}
\end{figure}

The non-linear interaction between the elastic stress and the velocity fields of a viscous fluid can cause an initially localized disturbance in the shape of a Gaussian pulse to split and spread spatially. Let us demonstrate this by employing a simplified one-dimensional viscoelastic fluid model. For that, consider the following pair of equations:
\begin{align}
\label{eq:1d_model_u}
\partial_t u_x &= -(u_x+U) \, \partial_x u_x + (b_x+B) \, \partial_x b_x + \nu \, \partial_{xx} u_x \\
\label{eq:1d_model_b}
\partial_t b_x &= -(u_x+U) \, \partial_x b_x + (b_x+B) \, \partial_x u_x - \lambda^{-1}  \, b_x
\end{align}
Here, $B$ represents a base elastic stress, and $U$ is the fluid's mean velocity component. In addition, $b_x$ and $u_x$ represent the time-dependent response of the elastic stress and the velocity to initial perturbations due to their non-linear interactions. This pair of equations is a toy model analogous to a one-dimensional version of the model derived in Refs.~\cite{Balkovsky2000, Fouxon2003}, which is founded upon the Oldroyd-B fluid model, assuming that the elastic stress tensor is uniaxial. For a case of zero elastic stress, $B=b_x=0$, eq.~\eqref{eq:1d_model_b} becomes irrelevant, and eq.~\eqref{eq:1d_model_u} reduces to the Burger's equation; in this case, an initial velocity pulse evolves into a pulse that is moderated by the viscous term (last term in eq.~\eqref{eq:1d_model_u}). However, if there is sufficiently strong base stress, $B$, the same initial velocity pulse can instead split into two pulses that travel away from each other.

A numerically solved example for the elastic pulse-splitting process is demonstrated in Fig.~\ref{fig:1d_model}, where $u_x(x)$ is plotted at four time values, and $b_x$ is shown as the shade of the curves. The initial condition is a Gaussian pulse for $u_x$ and $b_x(x)=0$ for the elastic stress. Following the initial velocity pulse, the elastic stress decreases (becomes negative) at the pulse's upstream edge and increases at its downstream edge, as shown for $t=0.3$. This initial generation of the elastic stress occurs through to the second term in the right-hand side of eq.~\eqref{eq:1d_model_b} which represents polymer stretching by the velocity gradient. Correspondingly, the gradient of the elastic stress develops a positive local maximum located between two negative local minima (a shape that resembles the Ricker wavelet function). At longer times, as demonstrated for $t=0.8$, the pulse splits into two peaks that are gradually growing apart. This splitting is driven by the second term on the right-hand side of eq.~\eqref{eq:1d_model_u} due to the three critical points of the elastic stress gradient, which represents the reaction of the elastic stress on the flow. Thus, Fig.~\ref{fig:1d_model} demonstrates that pulse splitting might occur in viscoelastic fluids as a result of non-linear interactions between the velocity and the elastic stress fields.

Whether the model results in pulse splitting of an initial single pulse sensitively depends on the details of the initial conditions and on the base stress levels. Specifically, the splitting is driven by the non-linear interaction of $u_x$ and $b_x$, through the second terms on the right side of eqs.~\eqref{eq:1d_model_u} and \eqref{eq:1d_model_b} so the rate for pulse-splitting grows with $|B|$. In addition, the generation of $b_x$ is driven by the polymer stretching term, which grows with the velocity gradient, $\partial_x u_x$, so the initiation of the process requires a sharp \sout{and} initial pulse with a sufficiently strong velocity gradient. Furthermore, the last term in eq.~\eqref{eq:1d_model_b} causes relaxation of the elastic stress with a timescale $\lambda$. Therefore, pulse-splitting can occur during the finite duration in which the perturbed fluid remains in the channel only if the relaxation time is sufficiently long and if the base elastic stress and the initial pulse strength are sufficiently high. Thus, the fact that the base stress in a channel flow grows with $\Wi$, suggests that pulse splitting may be observed in experiments at sufficiently high Weissenberg numbers, with fluids that have a sufficiently long relaxation timescale, and with sufficiently strong perturbations.

% ==========================================================
%                          Methods:
% ==========================================================

\section{Experimental methods}

\begin{figure}
	\centering
	\subfloat[]{\label{fig:setup}
		\includegraphics[width=8.5cm]{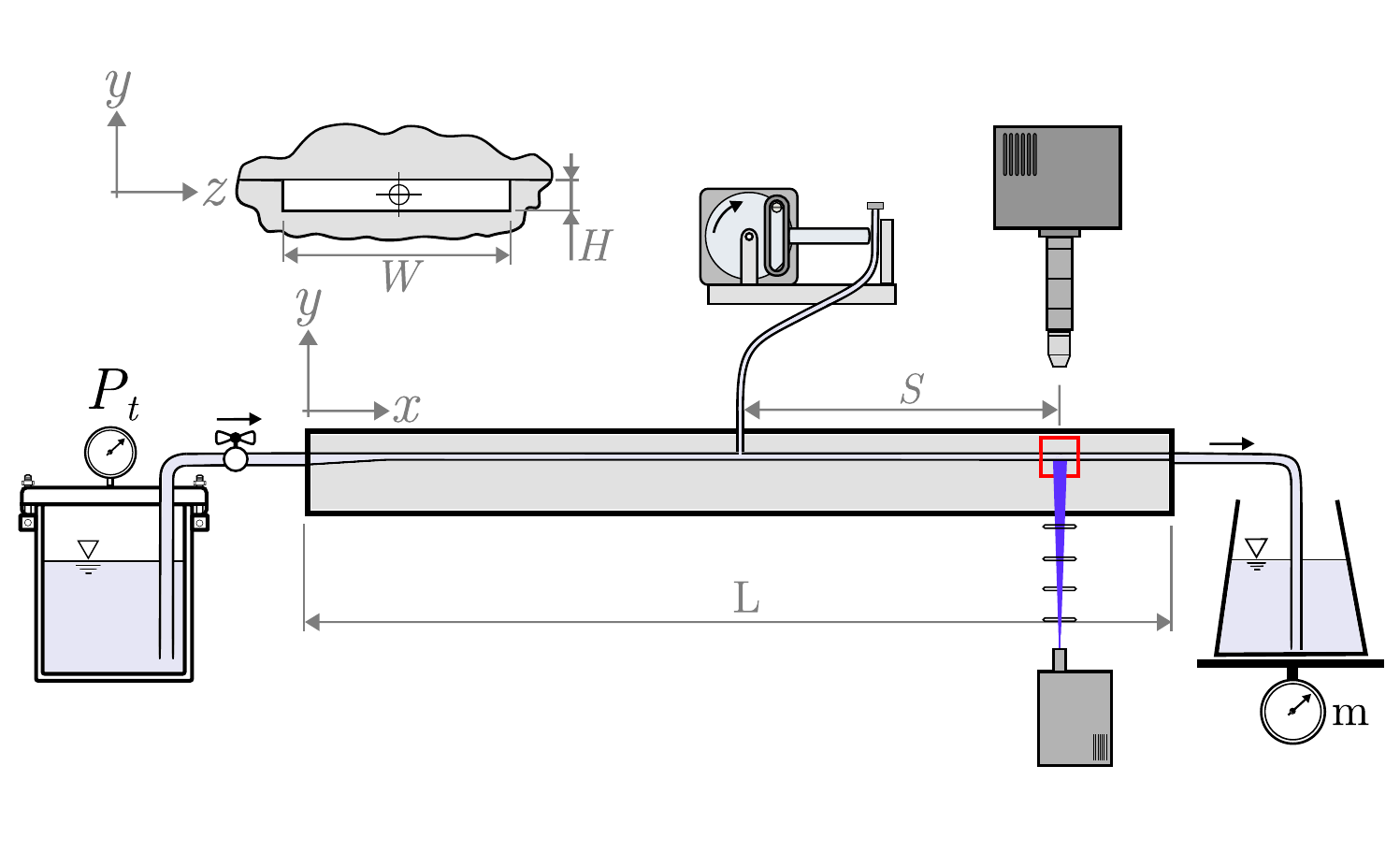}}
	\hfill
	\subfloat[]{\label{fig:pulse_example}
		\includegraphics[width=7.cm]{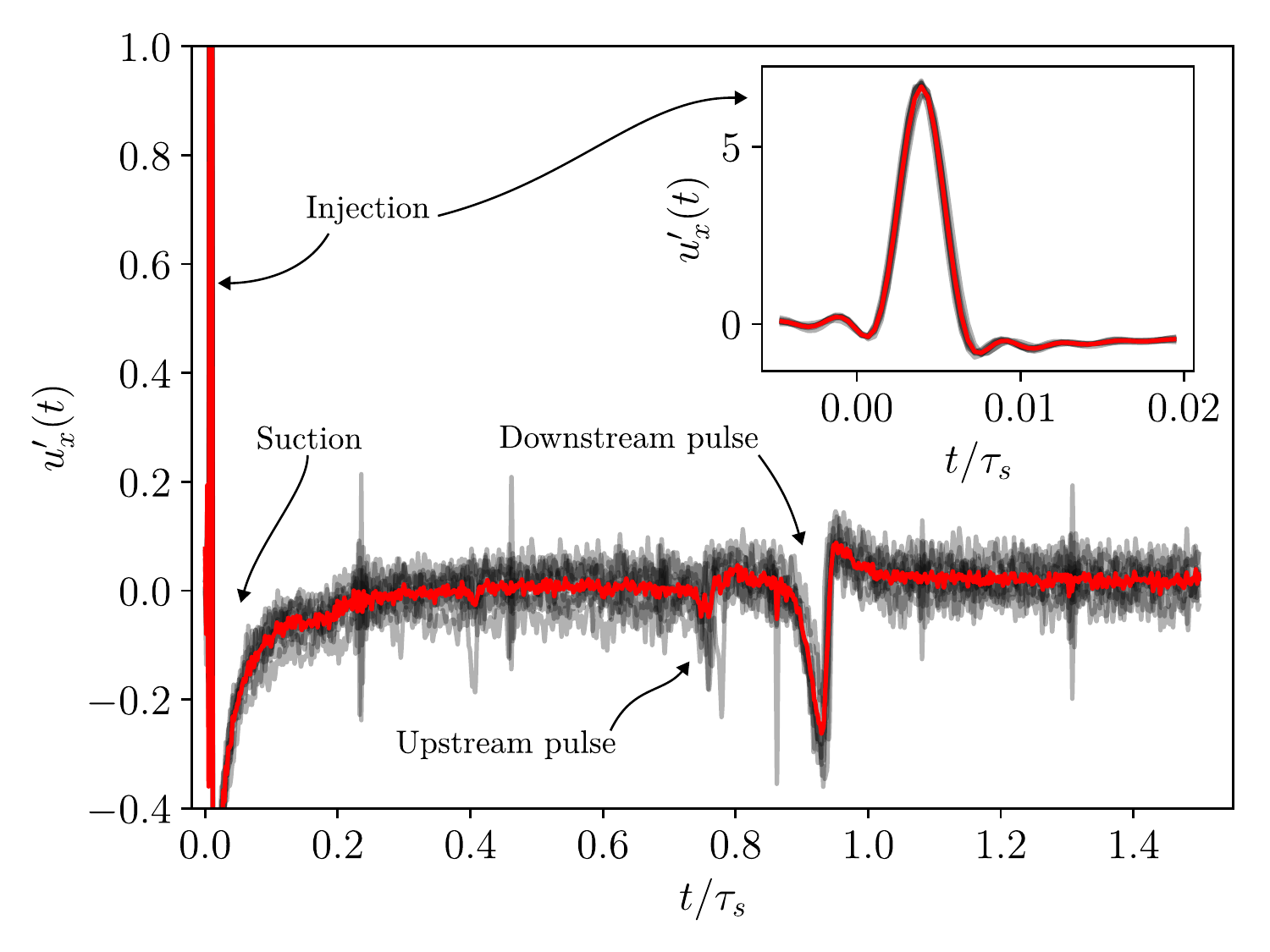}}
	\caption{(a) A schematic sketch of the experimental setup used. (b) Streamwise velocity time series taken at the center of the channel, $z=0$, that demonstrate the flow responce to perturbation at Weissenberg number lower than the critical value for growth of perturbaitons. Black lines show eight individual velocity time series after a perturbation event, and a red line shows their corresponding ensemble average; the inset focuses on the time of the injection. The flow parameters are $\Wi=166$, $\De=2.92$, taken at $S=486H$. }
\end{figure}

We study pulse-splitting experimentally by utilizing a long and narrow channel, with dimensions $750\times 3.5 \times 0.5$ mm in length, width, and height, as shown in Fig.~\ref{fig:setup}. The channels inlet was carefully smoothed and tapered to reduce unwanted perturbation that might trigger an instability~\cite{Shnapp2021}. As a working fluid we used a viscous solvent, comprised of 44\% Sucrose, 22\% D-Sorbitol, 1\% Sodium Chloride (\textit{Sigma Aldrich}), to which we added long polymers, Polyacrylamide (PAAm, $\mathrm{Mw}=18\times 10^6 Da$, (\textit{Polysciences inc}), at concentration $c=230$ ppm~\cite{Liu2009}. The solution properties are $\rho = 1320$ $\mathrm{Kg\,m^{-3}}$, $\eta_s=0.093 \, \mathrm{Pa\, s}$, and $\eta = \eta_p + \eta_s = 0.125 \, \mathrm{Pa\, s}$ are the solution density, the solvent's and the total solution's viscosity, respectively. The longest polymer solution relaxation time is $\lambda=12.1 \,[\mathrm{s}]$ based on the measurements of Ref.~\cite{Liu2009}. The viscous fluid was driven through the channel by using a pressurized Nitrogen tank, reaching up to roughly 80 psi.

We used a balance (BPS-1000-C2-V2, MRC) to measure the mass discharge rate $\Delta m/\Delta t$, so the mean flow velocity is calculated as $U=\frac{\Delta m}{\Delta t} / (\rho \, H \, W)$. Furthermore, we used the Particle Image Velocimetry (PIV) method to measure the flow velocity field over sections of the channel's central plane, $y=0$. The PIV measurements used a high-speed digital camera ((Photron FASTCAM Mini UX100), fitted with a 4X microscope objective (Nikon), an external function generator to trigger the camera, and 3.2 $\mu m$ fluorescent tracers particles. This setup provided measurement windows covering the full channel width ($W=7H$) and streamwise segments of up to $8.75H$ in length.

We perturbed the flow in localized events by executing fast injection and suction of fluid into and out of the channel. Perturbations were conducted through a small hole, $d_{\mathrm{p}} = 0.5$ mm in diameter, located at the center of one of the channel's vertical walls. We connected the small hole to a flexible tube to produce each injection-suction event. The tube, filled with fluid completely, was briefly squeezed and released using a PC-controlled stepper motor, producing the desired injection-suction perturbations. Each perturbation event used a fixed volume of fluid, $V_{\mathrm{p}}$, corresponding to a 0.75 mm long channel section: $V_{\mathrm{p}} = H\cdot W \cdot 0.75$ mm$^3$. Notably, the perturbation events did not alter the time-averaged $\Wi$ or $\Re$ numbers because the same amount of fluid injected was immediately extracted through suction. To control the amplitude of the perturbations, we changed the rate of injection, $t_{\mathrm{p}}$, which determines the velocity of the injected fluid. Thus, we characterize the perturbation amplitude by the dimensionless number: $\Wi_{\mathrm{p}} = \frac{V_{\mathrm{p}}}{\pi \frac{1}{4} d_{\mathrm{p}}^2 \, t_{\mathrm{p}}} \cdot \frac{\lambda}{d_p}$, which highlights the pulse-splitting requirement to produce polymer stretching. Furthermore, we conducted PIV measurements at several locations downstream from the injection hole to investigate the perturbed fluid's evolution. Denoting by $S$ the downstream distance from the hole, we conducted measurements at various distances in the range $S \in (110H, 582H)$. The time it takes the perturbed fluid packet to reach the measurement location, the transit time due to advection by $U$, is denoted $\tau_s=S/U$. Using $\tau_s$ we define the inverse Debora number as $\De^{-1} = \frac{\tau_s}{\lambda}$. Overall, our results span the three dimensional phase space: $\Wi$--$\Wi_{\mathrm{p}}$--$\De^{-1}$.

\begin{figure}
	\centering
	\subfloat[]{\label{fig:hi_wi}
		\includegraphics[width=7.5cm]{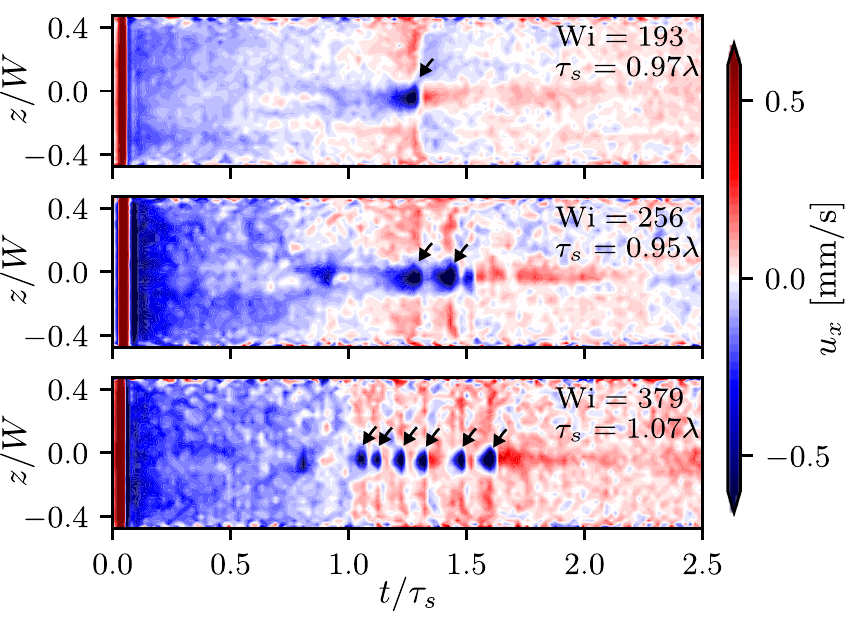}}
	\hfill
	\subfloat[]{\label{fig:count_histogram}
		\includegraphics[width=7.5cm]{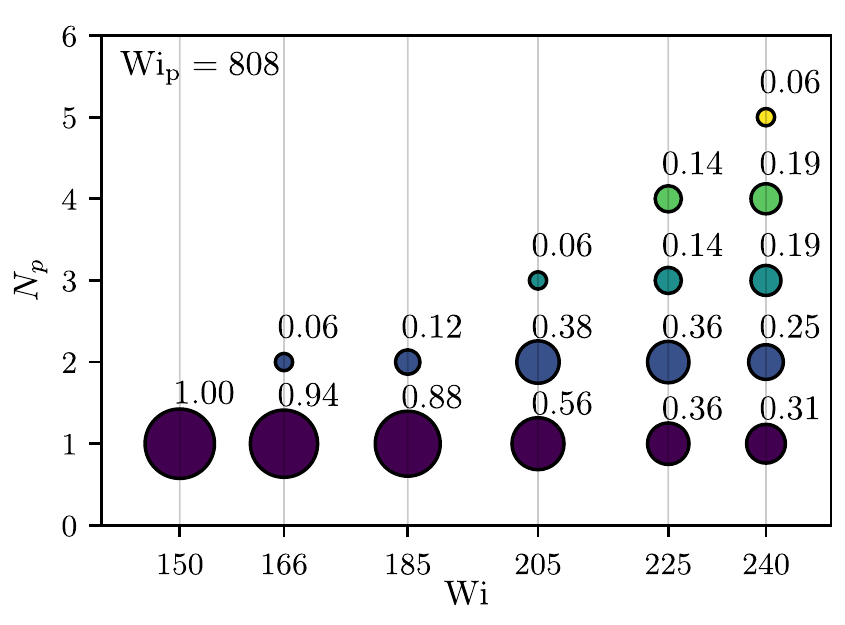}}
	\caption{(a) Space-time diagrams that demonstrate the streamwise velocity fluctuations following localized perturbations at Weissenberg numbers close to and above the onset of pulse-splitting. The flow is represented through contour plots of streamwise velocity fluctuations. The flow is perturbed at $t=0$, and at approximately $t\approx \tau_s$ numerous pulses (1, 2, or 6 depending on $\Wi$) can be identified as localized low velocity regions. Data is shown from three cases of roughly constant $\De^{-1}$ and three $\Wi$ values. (b) A series of probability distributions for the number of downstream pulses counted in repetitions of the experiment, $N_p$. Each column of circles corresponds to a fixed $\Wi$ case with a total of six cases, where $\Wi_{\mathrm{p}}=808$ is fixed. The area of each circle and the numbers printed correspond to the probability for observing $N_p$ for each $\Wi$ value. }
\end{figure}

% =========================================================
%                        Results 
% =========================================================

\section{Results}

In Fig.~\ref{fig:pulse_example}, we demonstrate how the perturbation and the flow's response is detected through our measurements by plotting time series of the streamwise velocity fluctuations ($u'_x=u_x-\overline{u_x}$) at the center of the channel, $z=y=0$. The measurement was made at $S=486 H$, far from the injection location, and for $\Wi=166$, which was not sufficiently high for pulse splitting to be observed. The injection is seen as a sharp "spike" at $t=0$, and the suction as a negative velocity excursion right after. These events cause the fluid in the channel to accelerate, decelerate, and accelerate back to its averaged value ($u'_x=0$). In addition, two distinct features can be recognized: two localized negative velocity pulses. There is a weaker pulse at $t\approx 0.75\tau_s$ and a much stronger pulse at $0.95\tau_s$. These localized features are the signature that we detected when the injected perturbed fluid had reached the measurement location (since $t\approx\tau_s=S/U$). Fig.~\ref{fig:pulse_example} demonstrates that this signature of the perturbed fluid was repeatable at this low $\Wi$ value since it shows eight repetitions of the measurements as black lines that are overlaid almost perfectly (up to experimental noise) by their ensemble average shown in red. Indeed, this signature signal, showing a single weak upstream pulse and a stronger downstream pulse, was repeatable in our measurements for sufficiently small $\Wi$. Furthermore, the pulses were observed for the full length of our channel and up to traveling times of $\tau_s=4.4\lambda$. Thus, these localized pulses constitute a stable state of the perturbed fluid after the perturbations. Notably, the same experiment was conducted with the Newtonian solvent at similar $\Re$ values, and no pulses could have been detected.

\begin{figure}
	\centering
	\subfloat[]{\label{fig:pulse_counts}
		\includegraphics[width=7.5cm]{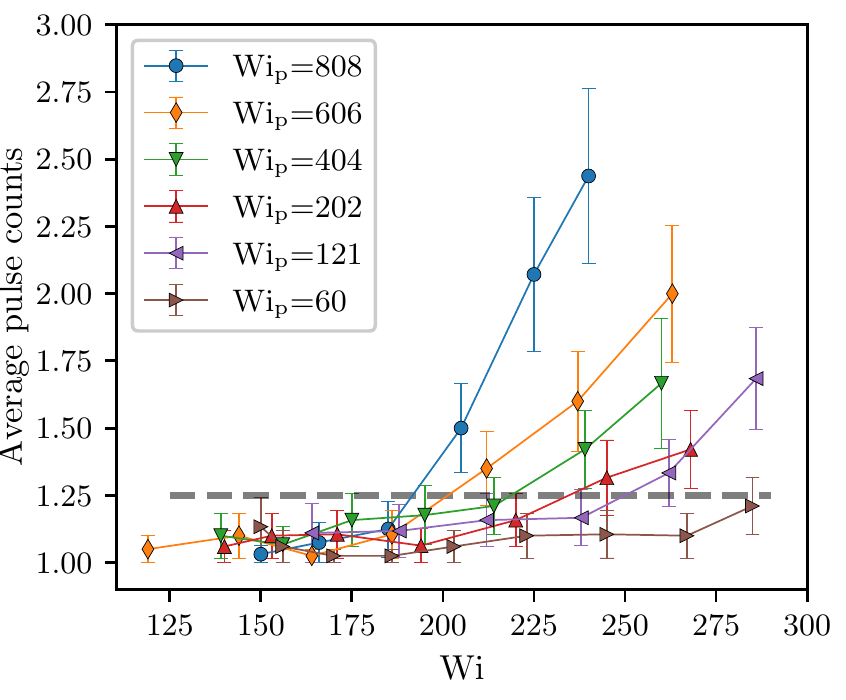}}
	\hfill
	\subfloat[]{\label{fig:critical_curve}
		\includegraphics[width=7.5cm]{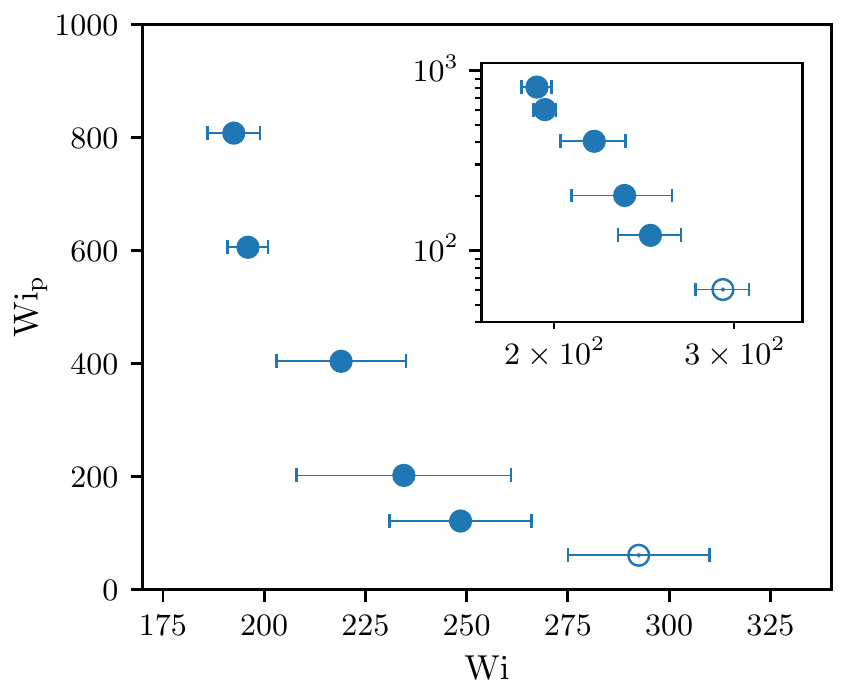}}
	\caption{(a) The average number of pulses, $\langle N_p \rangle$, observed at a distance of $S=486H$ downstream of the pulse injection location, shown as a function of the Weissenberg number for six levels of the pulse strength $\Wi_{\mathrm{p}}$. (b) The pulse strength, $\Wi_{\mathrm{p}}$, required to obtain the critical Weissenberg number $\Wi_c$, essentially showing the relation between the critical value for the transition and the perturbation strength. The inset shows the same data in log-log coordinates. The data and the error bars were calculated based on a linear interpolation of the data in (a).}
\end{figure}

At higher $\Wi$, the response to perturbations changed since we observed groups of several pulses instead of the single downstream pulse. Examples for observations with multiple pulses are shown in Fig.~\ref{fig:hi_wi} through space-time plots for a roughly fixed value of $\De^{-1}\approx 1$ and for three Weissenberg number values, $\Wi=193$, $256$ and $379$; the pulses are highlighted with small arrows. For the $\Wi=193$ case, a single downstream pulse, similar to Fig.~\ref{fig:pulse_example}, is observed at $t\approx1.25\tau_s$. The pulse is localized at the center of the channel, $z=0$, directly downstream to the injection hole. For the $\Wi=256$ case, at least two downstream pulses are identified, and a third weaker one may be just evolving. Moreover, six pulses can be identified for the $\Wi=379$ case. Therefore, we associate the observation that as $\Wi$ is increased, multiple pulses can be observed as a transition to a new state that can support pulse-splitting.

To investigate the transition to the new pulse-splitting state, we conducted measurements at various values of the empirical parameters with numerous repetitions for each case. The results revealed that the response to our perturbations in the pulse-splitting regime was stochastic. Specifically, the number of downstream pulses observed varied seemingly at random from one experimental repetition to another even at fixed $\Wi$, $\De^{-1}$, and $\Wi_{\mathrm{p}}$. Therefore, we denote the number of pulses observed after an injection event as $N_p$ and estimate the dependence of its statistics on the control parameters at a fixed location in the channel, $S=483H$, while varying $\Wi$ and $\Wi_{\mathrm{p}}$. For each value of the parameters, we used between 15 and 20 repetitions and gathered statistics of $N_p$. As an example, we show distributions of $N_p$ measurements for six values of $\Wi$ in Fig.~\ref{fig:count_histogram} for a fixed perturbation strength $\Wi_p = 808$. While for $\Wi=150$ only $N_p=1$ is observed in all repetitions, for $\Wi=240$ the observed pulse count varied in the range $ N_p \in ( 1 , 5 )$. This observation demonstrates that the pulse-splitting regime in our experiment was associated with a stochastic response to our perturbations.

Due to the observed stochasticity, we characterize the transition to the pulse-splitting state through the ensemble average $\langle N_p \rangle$. The dependence of $\langle N_p \rangle$ on $\Wi$ is shown in Fig.~\ref{fig:pulse_counts} for six values of the perturbation strength, spanning slightly more than an order of magnitude of $\Wi_{\mathrm{p}}$. For fixed $\Wi_{\mathrm{p}}$ values the average pulse count, $\langle N_p \rangle$, grows with $\Wi$ above some critical value. Furthermore, for weaker perturbation strengths, $\langle N_p \rangle$ grows slower, and the transition seems to occur at higher values of $\Wi$. Thus, we define an empirical threshold of $\langle N_p \rangle = 1.25 $ to mark the transition to the pulse-splitting regime (shown as a dashed line in Fig.~\ref{fig:pulse_counts}), and denote the critical Weissenberg number for this crossing as $\Wi_c$. As seen in Fig.~\ref{fig:pulse_counts}, $\Wi_c$ changes with the strength of the perturbation. Therefore, we plot $\Wi_c$ vs. $\Wi_p$ in linear scales in Fig.~\ref{fig:critical_curve}, showing the same data in log-log scales in the inset. Indeed, the perturbation strength value needed for the pulse-splitting transition decreases quickly with $\Wi$; roughly an order of magnitude reduction in $\Wi_p$ leads to a relatively modest change, only about 50\%, in $\Wi_c$. Due to the experimental uncertainty, a clear scaling is hard to obtain from the current data.

% ==========================================================
%                          Conclusions:
% ==========================================================

\section{Discussion and conclusions}

We report the first experimental analysis of the viscoelastic channel flow response to localized perturbation for $\Re\ll1$ and $\Wi\gg 1$. Specifically, a simplified model suggests that initially localized velocity pulses may undergo splitting due to the non-linear coupling of the velocity, the elastic stress, and their gradients. According to the model, splitting might be observed in the finite channel length for sufficiently high base elastic stress and perturbation strengths. Furthermore, pulse-splitting is confirmed experimentally in a long channel flow. Indeed, we observe a transition to a pulse-splitting regime at sufficiently high $\Wi$ and perturbation strengths, where multiple pulses are detected (up to six in this experiment) far away from the perturbation location, even though the same single perturbation event is produced.

The observation that by increasing $\Wi$ a localized perturbation may split agrees with the simplified model since the elastic stress in viscoelastic channel flow grows quadratically with $\Wi$ for the Oldroyd-B model of polymer~\cite{Morozov2007}. Therefore, as $\Wi$ increases in the experiment, $B$ grows as well, thus allowing pulse-splitting to be observed. In addition, we observe that the larger the perturbation strength, the smaller the Weissenberg number at which the pulse-splitting occurs, which agrees with the model as well.

While the simplified model is deterministic, the pulse-splitting regime in our experiment was stochastic since the number of pulses observed was random at fixed parameter values. Moreover, only up to one pulse-splitting event was observed in the model for each case, whereas multiple events (up to 5 splittings) were observed in the experiment after single perturbations. A plausible explanation for the observed stochasticity may be that the base elastic stress in the channel fluctuates in time due to seemingly minor imperfections in the experimental apparatus; in particular, the presence of the injection cavity was seen to generate an elastic instability in the region close the cavity, up to about $S=200H$~\cite{Shnapp2021}. Then, a pulse-splitting event might occur when the elastic stress fluctuations become sufficiently strong. Notably, the current observations of the pulse-splitting regime were intentionally conducted very far downstream from the hole (e.g., $S=483H$ in Figs.~\ref{fig:count_histogram}, \ref{fig:pulse_counts}, \ref{fig:critical_curve}) at locations out of the direct influence of the hole-related instability. In addition, although the simplified model we used recovers the fundamental splitting process itself, the flow in the channel is intrinsically three-dimensional and thus much more complex than what a one-dimensional model could achieve. Indeed, a more elaborate modeling effort is required to fully resolve the observed transition's details.

Our results indicate a new possible scenario for the growth of perturbations through an intrinsically elastic process. This observation is important since it suggests a plausible new route for the elastic instability and elastic turbulence to develop in the viscoelastic channel flows. In addition, our new approach to the problem opens a new direction for continued research on this topic.

%\begin{acknowledgements}
We are grateful to Guy Han, Rostyslav Baron, and Gershon Elazar for their assistance with the experimental setup. This work was partially supported by grants from the Israel Science Foundation (ISF; grant \#882/15 and grant \#784/19) and the Binational USA-Israel Foundation (BSF; grant \#2016145). RS is grateful for the financial support provided by the Clore Israel Foundation.
%\end{acknowledgements}

\bibliographystyle{abbrv}
\bibliography{./bib}

\clearpage

\end{document}